%% file: main.tex
\PassOptionsToPackage{hypertexnames=false}{hyperref} % get rid of warnings
\documentclass[copyright,creativecommons]{eptcs}

\usepackage[utf8]{inputenc}
\usepackage[T1]{fontenc}
\usepackage{newunicodechar}
\input{unicodechars}

\providecommand{\event}{ThEdu'18 Post-Proceedings} % Name of the event you are submitting to
\usepackage{underscore}           % Only needed if you use pdflatex.

\usepackage{csquotes}
\usepackage{verbatim}
\usepackage[nounderscore]{syntax}
\usepackage{graphicx}
\usepackage{amsmath}
\usepackage{amsthm}
\usepackage{amssymb}
\usepackage{mathbbol}
\DeclareSymbolFontAlphabet{\mathbb}{AMSb}

\makeatletter
\def\verbatim@font{\scriptsize\ttfamily}
\makeatother
\input{commands}

\title{Theorem and Algorithm Checking for \\ 
Courses on Logic and Formal Methods\thanks{Supported by
the Johannes Kepler University Linz, Linz Institute of Technology (LIT), 
Project LOGTECHEDU \enquote{Logic Technology for Computer Science Education}
and by the OEAD WTZ project SK 14/2018 SemTech.}}

\author{Wolfgang Schreiner
\institute{Research Institute for Symbolic Computation (RISC) \&
Linz Institute of Technology (LIT) \\
Johannes Kepler University, Linz, Austria}
\email{Wolfgang.Schreiner@risc.jku.at}}

\def\titlerunning{Theorem and Algorithm Checking for Courses on
Logic and Formal Methods}
\def\authorrunning{Wolfgang Schreiner}

\begin{document}
\maketitle

\begin{abstract}

The RISC Algorithm Language (RISCAL) is a language for the formal modeling of
theories and algorithms. A RISCAL specification describes an infinite class of
models each of which has finite size; this allows to fully automatically check
in such a model the validity of all theorems and the correctness of all
algorithms. RISCAL thus enables us to quickly verify/falsify the specific truth
of propositions in sample instances of a model class before attempting to prove
their general truth in the whole class: the first can be achieved in a fully
automatic way while the second typically requires our assistance. RISCAL has
been mainly developed for educational purposes. To this end this paper reports
on some new enhancements of the tool: the automatic generation of checkable
verification conditions from algorithms, the visualization of the execution of
procedures and the evaluation of formulas illustrating the computation of their
results, and the generation of Web-based student exercises and assignments from
RISCAL specifications. Furthermore, we report on our first experience with
RISCAL in the teaching of courses on logic and formal methods and on further
plans to use this tool to enhance formal education.

\end{abstract}

\section{Introduction}
\label{sect:introduction}

Teaching courses that rely on logic-based formalisms (such as the
specification and verification of computer programs) is hampered by the
difficulty of conveying to students the meaning of logic formulas
respectively the subtly different meanings of multiple variants of logic
formulations. This is especially the case if the effects of such
formulations are only exhibited by the consequent provability of
propositions (such as verification conditions whose validity implies the
correctness of algorithms). The use of a tool that automates a proving
calculus is here typically of little help: first, since such a tool (in a
sufficiently rich logic) only provides semi-decisions, a conjecture may
still be correct, even if it cannot be proved by the tool; second, even if a
conjecture is indeed wrong, a failed proof attempt typically does not shed
much light on the core reason. This is the more deplorable since in practice
(as well in education as in research) most time is actually spent in trying
to prove conjectures that do \emph{not} hold. This is also the fundamental
limitation on the educational use of tools for deductive program
verification, be they \enquote{interactive}, \enquote{automatic}, or
\enquote{auto-active}; prominent examples of such tools are the interactive
KeY verifier~\cite{Ahrendt2016}, the verification platform
Why3~\cite{Bobot2011} with various interactive and automatic backends, the
OpenJML tool with various SMT solvers as automatic backends~\cite{Cok2011},
and the auto-active systems Spec\#~\cite{Barnett2004} and
Dafny~\cite{Leino2010}, both based on the Boogie backend.

One attempt to overcome these problems is to use reasoning tools that are based
on model checking~\cite{Clarke2018} rather than proving: first, such tools
generally provide full decisions (if the violation of a conjecture is reported,
the conjecture indeed does not hold); second, if a conjecture does not hold,
such tools provide a counter-example that may help to understand the underlying
reason. The downside of this ability of model checkers is the necessity to
restrict the domain of investigation to models of finite size whose properties
are consequently decidable. Furthermore, since model checkers have arisen in the
context of the verification of hardware/software systems, they typically only
support specification languages of limited expressiveness, usually based on a
quantifier-free logic (propositional logic, propositional linear temporal logic,
or some other logic where the satisfiability problem is decidable); if also
quantifiers are supported, checkers typically apply heuristics and thus loose
decidability, yielding the same disadvantages that plague automated provers.
This makes these tools less suitable for the formal education of students of
computer science or mathematics.

The RISC Algorithm Language (RISCAL)~\cite{Schreiner2017,RISCAL} is an attempt
to make model checking an attractive option also for such educational scenarios.
RISCAL does not at all compromise the expressiveness of its specification
language: it supports full (first-order) predicate logic on top of a rich basis
of types (integers, tuples, records, arrays, sets, recursive types) with
corresponding operations. To make the resulting theory still decidable, the only
restriction is that all types have bounds that make the model finite: e.g.,
\texttt{$\mathbb{Z}[N,M]$} denotes the type of all integers in interval $[N,M]$
while \texttt{Array[N,T]} denotes the type of all arrays of length $N$ whose
values have type $T$. However, these type bounds may be unspecified constants,
thus each RISCAL specification denotes an infinite class of models where each
instance of this class is finite. By choosing some values for these constants we
may pick an instance of this class and check all its properties that are
expressible in the logic. The basic technique to check the validity of a formula
(and in general of every phrase in RISCAL such as terms and commands) is to
evaluate its semantics; for instance to evaluate a formula \texttt{($\forall
x$:$T$.\,$F$)} we enumerate all values of type~$T$, bind variable $x$ to each of
these values in turn and evaluate the formula body $F$ in this binding. Since
the language also supports constructs such as \texttt{(choose~$x$:$T$.\,$F$)}
which denotes an arbitrary value $x$ of type $T$ that satisfies formula~$F$, the
semantics of RISCAL is in general multi-valued: if we select for the model
checker a \enquote{nondeterministic} evaluation mode, all possible outcomes are
computed (in a lazy fashion); in its (faster) \enquote{deterministic} mode, only
one of the outcomes is considered.

In predecessor publications we have already described the basic capabilities of
RISCAL~\cite{Schreiner2018a,Schreiner2018d}. The purpose of this paper is to
outline some recent enhancements that target the usefulness of RISCAL in
educational scenarios: the automatic generation of checkable verification
conditions, the visualization of procedure execution and formula evaluation, and
the generation of Web-based student exercises respectively assignments from
RISCAL specifications. Furthermore, we give first reports on our experience of
the use of RISCAL in the classroom and sketch future plans.

RISCAL is not the first attempt to apply model checking technology in formal
modeling: various other modeling languages support checking respectively
counterexample generation~\cite{ABZ2016}, differing from RISCAL mainly in the
application domain of the language and/or the exhaustiveness of the analysis.
The development of RISCAL has been very much inspired by the TLA Toolbox which
incorporates a model checker for a subset of TLA+~\cite{Lamport2002}; also the
use of unspecified constants as model bounds has been inspired by this toolkit.
However, TLA+ is untyped and actually also supports models of infinite size;
thus the toolkit does not really represent a reliable decision procedure for
TLA+ specifications.

The specification language Alloy~\cite{Jackson2011} is based on a relational
logic designed for modeling the states of abstract systems and the executions of
such systems; given bounds for state sizes and execution lengths, the Alloy
Analyzer finds all executions leading to states satisfying/violating given
properties and thus indeed represents a decision procedure. However, the
specification language of Alloy is quite system-oriented which makes it not very
attractive for describing general mathematical domains and
algorithms~\cite{Ritirc2016}. On the other hand, the counterexample generator
Nitpick~\cite{Blanchette2010} for the proof assistant Isabelle/HOL supports
classical predicate logic but may (due to the presence of unbounded quantifiers)
fail to find counterexamples; in an \enquote{unsound mode} quantifiers are
artificially bounded, but then invalid counterexamples may be reported.

While Alloy and Nitpick are based on SAT-solving techniques, RISCAL's
implementation is actually closer to that of the test case generator
Smallcheck~\cite{Runciman2008}. This system generates for the parameters of a
Haskell function in an exhaustive way argument values up to a certain bound; the
results of the corresponding function applications may be checked against
properties expressed in first-order logic (encoded as executable higher-order
Haskell functions). However, unlike RISCAL, the value generation bound is
specified by global constants rather than by the types of parameters and
quantified variables such that separate mechanisms have to be used to restrict
searches for counterexamples.

The remainder of this paper is organized as follows: To make the presentation
self-contained, we give in Section~\ref{sect:riscal} (based on previously
published material) a short overview on the core functionality of RISCAL. In
Section~\ref{sect:features} we discuss new (not yet formally published) features
of RISCAL. In Section~\ref{sect:education} we discuss our experience with and
plans for the use of RISCAL in education. Section~\ref{sect:conclusions}
presents our conclusions and sketches ideas for the future.

\section{The RISCAL System}
\label{sect:riscal}

\begin{figure}
\centering
\includegraphics[width=0.8\textwidth]{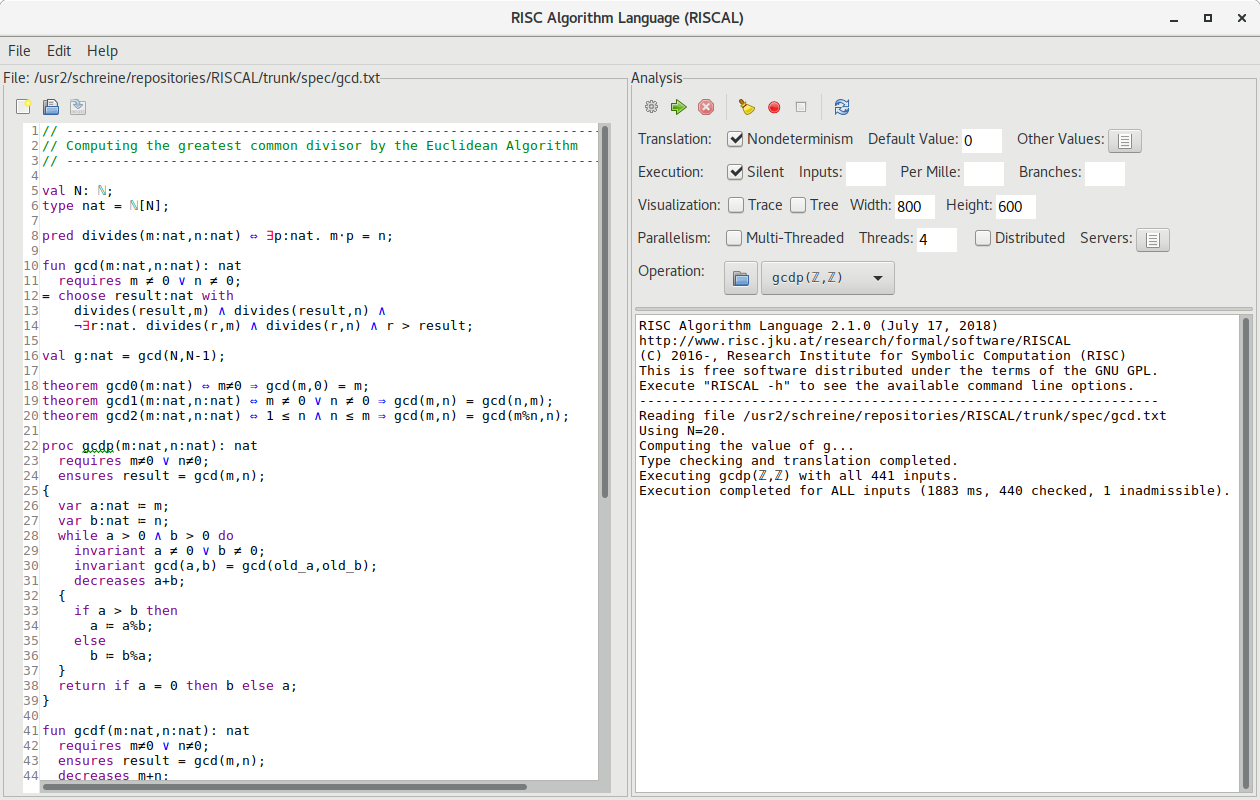}
\caption{The RISCAL System}
\label{fig:RISCAL}
\end{figure}

To make this paper self-contained, we give a short summary overview on RISCAL
with material taken from~\cite{Schreiner2018a,Schreiner2018d}; a more detailed
tutorial on the use of the software and the complete reference of its
specification language can be found in the manual~\cite{Schreiner2017}.

Figure~\ref{fig:RISCAL} depicts the user interface of the RISCAL software with
an editor frame for RISCAL specifications on the left and the control widgets
and output frame of the checker on the right. The RISCAL specification language
is based on a statically typed variant of first order predicate logic. On the
one hand, it allows to develop mathematical theories such as that of the
greatest common divisor depicted at the top of Figure~\ref{fig:gcd}; on the
other hand, it also enables the specification of algorithms such as Euclid's
algorithm depicted in the same figure below. The lexical syntax of the language
includes Unicode characters for common mathematical symbols; these may be
entered in the RISCAL editor via ASCII shortcuts.

\begin{figure}[t]
\begin{xverbatim}
val N: ℕ; type nat = ℕ[N];

pred divides(m:nat,n:nat) ⇔ ∃p:nat. m⋅p = n;
fun gcd(m:nat,n:nat): nat
  requires m ≠ 0 ∨ n ≠ 0;
= choose result:nat with
    divides(result,m) ∧ divides(result,n) ∧
    ¬∃r:nat. divides(r,m) ∧ divides(r,n) ∧ r > result;

theorem gcd0(m:nat) ⇔ m≠0 ⇒ gcd(m,0) = m;
theorem gcd1(m:nat,n:nat) ⇔ m ≠ 0 ∨ n ≠ 0 ⇒ gcd(m,n) = gcd(n,m);
theorem gcd2(m:nat,n:nat) ⇔ 1 ≤ n ∧ n ≤ m ⇒ gcd(m,n) = gcd(m%n,n);

proc gcdp(m:nat,n:nat): nat
  requires m≠0 ∨ n≠0;
  ensures result = gcd(m,n);
{
  var a:nat ≔ m; var b:nat ≔ n;
  while a > 0 ∧ b > 0 do
    invariant a ≠ 0 ∨ b ≠ 0;
    invariant gcd(a,b) = gcd(old_a,old_b);
    decreases a+b;
  {
    if a > b then a ≔ a%b; else b ≔ b%a;
  }
  return if a = 0 then b else a;
}
\end{xverbatim}
\caption{Euclid's Algorithm in RISCAL}
\label{fig:gcd}
\end{figure}

The specification listed in Figure~\ref{fig:gcd} introduces the mathematical
theory of the greatest common divisor and its computation by the Euclidean
algorithm; this theory is restricted to the domain of all natural numbers less
than equal \texttt{$N$}. In particular, it defines a predicate
\texttt{divides($m$,$n$)} which denotes $m|n$ ($m$ divides $n$) and subsequently
(by an implicit definition) a function \texttt{gcd(m,n)} which denotes the
greatest common divisor of $m$ and $n$. The theorems \texttt{gcd0($m$)},
\texttt{gcd1($m$,$n$)}, and \texttt{gcd2($m$,$n$)} describe the essential
mathematical propositions on which the correctness proof of the Euclidean
algorithm is based. The procedure \texttt{gcdp($m$,$n$)} embeds an iterative
implementation of the Euclidean algorithm. Its contract specified by the clauses
\texttt{requires} and \texttt{ensures} states that the procedure behaves exactly
as the implicitly defined function; the loop annotations \texttt{invariant} and
\texttt{decreases} describe essential knowledge for proving the total
correctness of the procedure.

For all parameterized entities (functions, predicates, theorems, procedures) the
menu \enquote{Operation} allows to select the entity; by pressing the
\enquote{Run} button \button{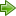}, the system generates (in a lazy fashion)
all possible combinations of parameter values that satisfy the precondition of
the operation, invokes the operation on each of these, and prints the
corresponding result values. If the selection box \enquote{Silent} is checked,
the output for each operation is suppressed; however, each execution still
checks the correctness of all annotations (preconditions, postconditions,
theorems, invariants, termination measures, and assertions). If the checking
thus completes without errors, we have validated that the operation satisfies
the specification for the domains determined by the current choices of the
domain bounds.

For instance, we can check the validity of theorem \texttt{gcd2}:
\begin{xverbatim}
Executing gcd2(ℤ,ℤ) with all 441 inputs.
Execution completed for ALL inputs (256 ms, 441 checked, 0 inadmissible).
\end{xverbatim}
This ensures that for all possible parameter values, the truth value of the
corresponding formula is \enquote{true}. Likewise, we can validate that the
procedure \texttt{gcdp} satisfies its specification:
\begin{xverbatim}
Executing gcdp(ℤ,ℤ) with all 441 inputs.
Execution completed for ALL inputs (933 ms, 440 checked, 1 inadmissible).
\end{xverbatim}
This check indeed evaluates the procedure specification and the embedded
loop annotations; here one input (that with $m=n=0$) is reported as
\enquote{inadmissible} which indicates that it violates the precondition of
the procedure; consequently, the procedure need not be checked with that
input.

\section{New Features of RISCAL}
\label{sect:features}

In this section, we report on new features of RISCAL relevant for its
application in the context of educational scenarios; these features have been
described in a series of technical
reports~\cite{Schreiner2017,Schreiner2018b,Schreiner2018c,Schreiner2018e}
but not yet been formally published before.

\subsection{Generation and Checking of Verification Conditions}
\label{sect:verification}

To further validate the correctness of algorithms (and as a prerequisite of
subsequent general correctness proofs), RISCAL provides since version 2.0 a
\emph{verification condition generator (VCG)}. This VCG produces from the
definition of an operation (including its specification and other formal
annotations) a set of verification conditions, i.e., theorems whose validity
implies that, for all arguments that satisfy the precondition of the operation,
its execution terminates and produces a result that satisfies its postcondition.
These theorems are plain RISCAL formulas; their validity (in a finite domain)
can be automatically checked within RISCAL itself.

This VCG is fundamentally based on Dijkstra's weakest precondition (wp)
calculus. In order to prove the correctness of a procedure
\begin{xverbatim}
proc p(x:T1):T2 requires P(x); ensures Q(x,result) { C; return r; }
\end{xverbatim}
we generate a theorem
\begin{xverbatim}
theorem _p_Corr(x:T1) requires P(x); ⇔ wp(C, let result = r in Q);
\end{xverbatim}
where $\mathit{wp}(C,Q)$ is a formula generated from the body $C$ of the
procedure and its postcondition~$Q$ (called \enquote{the weakest precondition of
$C$ with respect to $Q$}). In a nutshell, the theorem states that every
argument~$x$ that satisfies the precondition~$P$ of the procedure also satisfies
that weakest precondition (however, as discussed below, we actually generate not
only one such big formula but a lot of smaller ones).

Generating and checking verification conditions in RISCAL serves a particular
purpose: it not only ensures that the algorithm works as expected (this has been
demonstrated already by checking the algorithm) but also establishes that all
annotations (specifications, loop invariants, termination terms) are strong
enough to verify the correctness of the algorithm by formal proof. If checking
the verification conditions succeeds, this indeed demonstrates that such a proof
is possible in the particular model (determined by the concrete values assigned
to the type bounds) in which the check has taken place. However, the success of
such a check gives also reason to believe (at least increases our confidence)
that such a proof may be possible for the infinite class of \emph{all} possible
models (i.e., for all infinitely many possible values of the type bounds). At
least, if a check fails, we know that there is no point in attempting such a
general proof before the error exhibited in the particular model has been fixed.

Our default assumption is that (due to some problem in the definition of the
operation respectively in its specification or annotations) a correctness
proof will fail. Therefore RISCAL tries to help the user to find the reason
of such a failure by the following strategy:
\begin{itemize}
\item Rather than generating a single big verification condition whose validity
ensures the overall correctness of the algorithm, RISCAL generates a lot of
smaller verification conditions each of which demonstrates some aspect of
correctness. Thus, if a particular condition fails, we do not only know that the
overall correctness of the algorithm cannot be established but we can focus on
that aspect of correctness expressed by the corresponding condition. 
\item Each verification condition is linked to those parts of the algorithm
from which the condition has been generated and that are therefore relevant
for its validity. These parts are visualized in the editor such that we
get a quick intuition on which parts of the algorithm we have to concentrate.
\end{itemize}

\begin{figure}
\begin{center}
\includegraphics[width=0.25\textwidth]{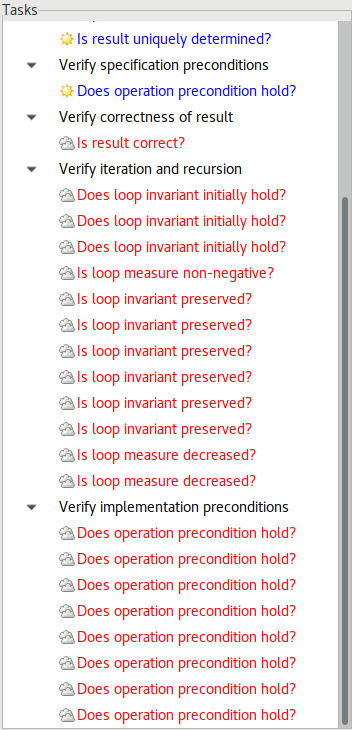}
\includegraphics[width=0.25\textwidth]{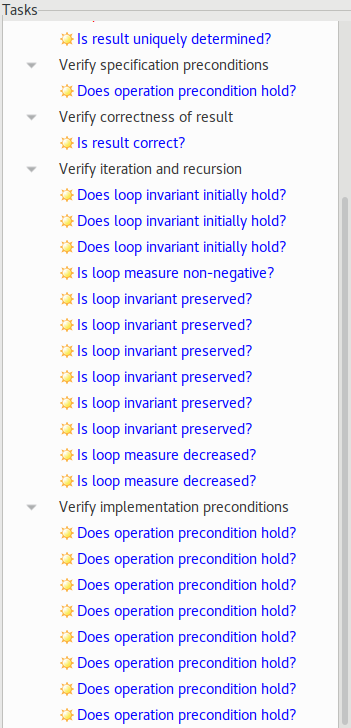}
\end{center}
\caption{The Correctness of the Algorithm}
\label{Correctness}
\end{figure}

In case of the greatest common divisor procedure \texttt{gcdp} that was sketched
in Section~\ref{sect:riscal}, RISCAL generates the verification conditions
listed in Figure~\ref{Correctness}. These conditions are initially indicated as
red tasks; having successfully checked them in RISCAL (by double clicking the
tasks or right-clicking the task and selecting in the pop-up menu the entry
\button{go-next} \enquote{Execute Task}), the tasks turn blue. These tasks check
whether the result is correct (whether the precondition of the procedure implies
the core of the weakest precondition of its body with respect to the specified
postcondition), whether the loop invariant initially holds, whether the loop
measure is non-negative, whether the loop invariant is preserved, whether the
loop measure is decreased, and whether the preconditions of the various
operations hold.

\begin{figure}
\begin{center}
\includegraphics[width=0.45\textwidth]{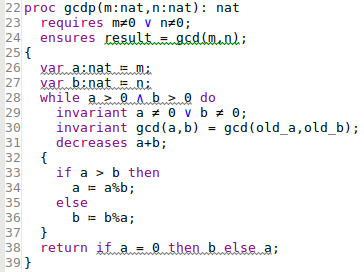}
\end{center}
\caption{Code Relevant to Verification Condition \enquote{Is Result Correct?}}
\label{GCD-Correctness}
\end{figure}

For instance, the task \enquote{Is the result correct?} checks the validity
of the following verification condition (whose definition can be displaced
	by clicking a corresponding entry in a popup menu):
\begin{xverbatim}
theorem _gcdp_5_CorrOp0(m:nat, n:nat)
requires (m ≠ 0) ∨ (n ≠ 0);
 ⇔ let a = m in (let b = n in 
   (letpar old_a = a, old_b = b in 
   (∀a:nat, b:nat. (((((a ≠ 0) ∨ (b ≠ 0)) ∧ 
   	               (gcd(a, b) = gcd(old_a, old_b))) ∧ 
   	               (¬((a > 0) ∧ (b > 0)))) ⇒ 
     (let result = if a = 0 then b else a in 
       (result = gcd(m, n)))))));
\end{xverbatim}
Furthermore, by selecting the condition with a single click, the editor 
displays the view shown in Figure~\ref{GCD-Correctness}. The postcondition to be
proved is underlined in green while grey underlines indicate those parts of the
program that contributed to the derivation of the weakest precondition (in
case of a loop, only the loop expression is underlined; actually, most of the
verification condition comes from the invariant of the loop). It should be noted
that the body of the loop does here play no role, because for the derivation
of the weakest precondition only the invariant of the loop is considered.
This condition (like all the other verification conditions) can be automatically
checked by a double mouse click.

\subsection{Visualization of Procedure Execution and Formula Evaluation}
\label{sect:visualization}

Since version 2.1.0 RISCAL provides both the visualization of execution
traces~\cite{Schreiner2018b} and of formula evaluation~\cite{Schreiner2018c}.
For this the graphical user interface depicts a row \enquote{Visualization} with
two (mutually exclusive) check box options \enquote{Trace} and \enquote{Tree}.

By selecting the option~\enquote{Trace} the execution of a procedure displays
the changes of the variable values by the commands in the body of a procedure;
for functions (including predicates and theorems) whose result is determined by
the evaluation of an expression the computation of the result is displayed as a
single step (except that also calls of other operations are displayed). This
mode of visualization is thus mainly useful for understanding the operational
behavior of a procedure. For instance, take the following RISCAL model
of the \enquote{Bubble sort} algorithm:
\begin{quote}\small
\begin{verbatim}
val N:ℕ; val M:ℕ;
type index = ℤ[-N,N]; type elem = ℤ[-M,M]; type array = Array[N, elem];

proc cswap(a:array, i:index, j:index): array
{
  var b:array = a;
  if b[i] > b[j] then {
    var x:elem ≔ b[i];
    b[i] ≔ b[j];
    b[j] ≔ x;
  }
  return b;
}

proc bubbleSort(a:array): array {
  var b:array = a;
  for var i:index ≔ 0; i < N-1; i ≔ i+1 do {
    for var j:index ≔ 0; j < N-i-1; j ≔ j+1 do
      b ≔ cswap(b,j,j+1);
  }
  return b;
}
\end{verbatim}
\end{quote}

\begin{figure}
\begin{center}
\includegraphics[width=0.99\textwidth]{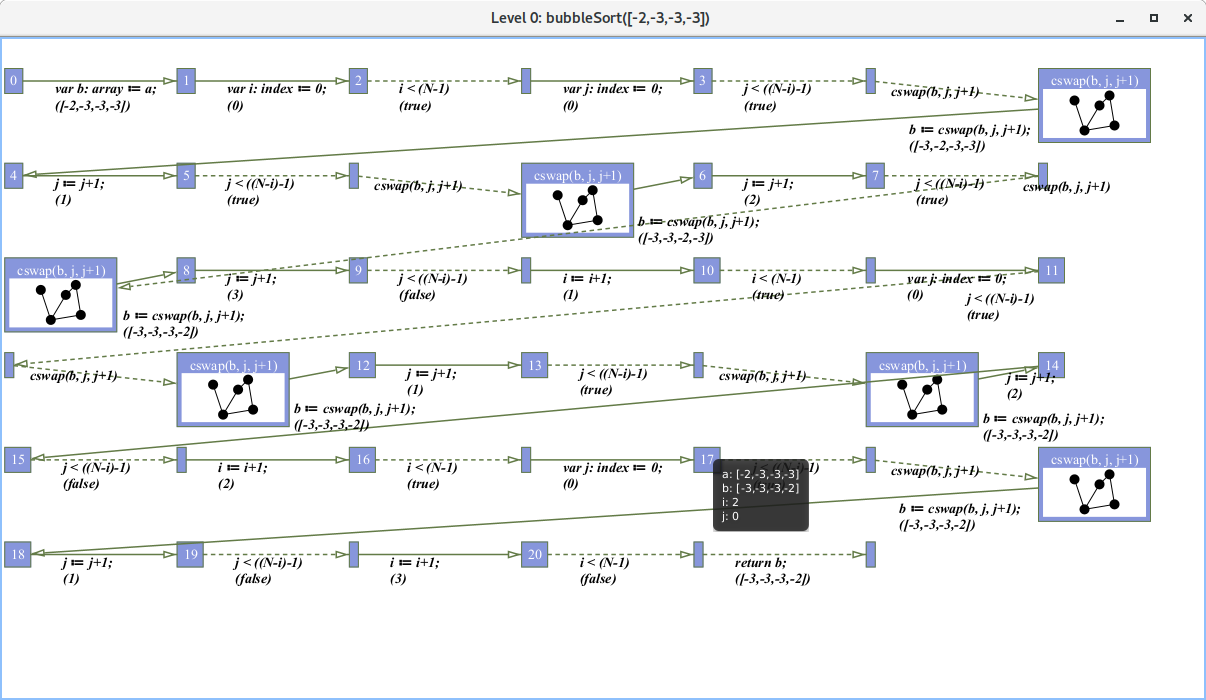} %\\[1em]
\end{center}
\caption{Visualization of Bubble Sort}
\label{fig:bubble}
\end{figure}

If we select the operation \texttt{bubbleSort} and set with the button
\enquote{Other Values} the parameters $N$ and $M$ to $4$ and $3$, respectively,
the second check of the procedure is performed with the argument array
$a=[-2,-3,-3,-3]$ upon which the window displayed in Figure~\ref{fig:bubble}
pops up. The window displays a directed graph (a linear sequence of nodes) that
are connected by directed edges (arrows) and that are laid out in a
two-dimensional manner from left to right and top to bottom. The numbered nodes
in this graph represent the sequence of states constructed by performing
assignments to the variables of the procedure; by hovering with the mouse
pointer over such a node, a small window pops up that displays the values of the
various variables in that state (see the node numbered 17 in the figure). The
nodes with a graph symbol represents the call of another operation, the header
of this window displays the call itself. By double-clicking on such a call node
which represents the application of an operation the content of the window is
modified to visualize the execution of that operation (moving to another level
of visualization). A double click on any empty part of the window moves the
display back to the previous level.

The option~\enquote{Tree} triggers the visualization of the evaluation of
logic formulas. As an example, we visualize in the following the evaluation of
the formula introduced by the following definition of predicate
\texttt{forallPexistsQFormula()}:
\begin{flushleft}
\small
\begin{verbatim}
  val N = 4;
  type T = ℕ[N];
  pred p(x:T) ⇔ x < N;
  pred q(x:T,y:T) ⇔ x+1 = y;
  pred forallPexistsQFormula() ⇔ ∀x:T. p(x) ⇒ ∃y:T. q(x,y);
\end{verbatim}
\end{flushleft}

\begin{figure}
\begin{center}
\includegraphics[width=0.8\textwidth]{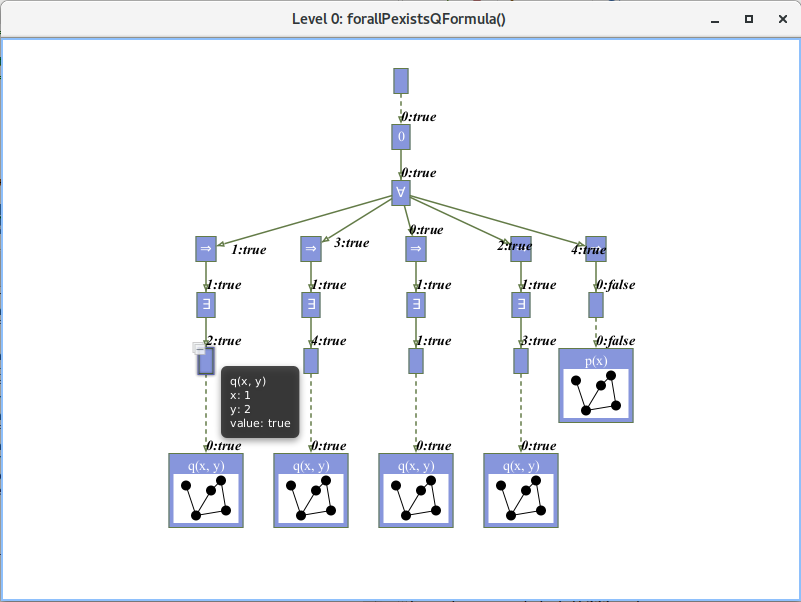}
\end{center}
\caption{The Visualization of the Evaluation of
$(\forall x.\ p(x) \Rightarrow \exists y.\ q(x,y))$}
\label{fig:tree}
\end{figure}

If this formula is evaluated, the window depicted in Figure~\ref{fig:tree} pops
up which displays the evaluation tree for that formula. The root of that tree
represents the evaluation of the predicate for all possible arguments. Numbered
nodes are children of the root node each of which represents the evaluation of
the predicate for one particular argument. Labeled nodes represent propositional
formulas or quantified formulas where the label denotes the formula's outermost
logical symbol (logical connective or quantifier). By hovering with the mouse
pointer over such a node, a small window pops up that displays the formula, the
values of its free variables (actually the values of all variables visible at
the occurrence of the formula), and the truth value of the formula. Atomic
formulas with application of user-defined predicates are displayed by a node
with a graph symbol. By double-clicking on such a node, the visualization
switches from the current formula to the formula defining the predicate.

Every evaluation tree is \emph{pruned} such that it only displays the
information necessary to understand how its truth value was derived. For
instance, if the truth value of a conjunction $(F_1 \wedge F_2)$ is
\enquote{false}, only the first subformula is displayed whose truth value is
\enquote{false}. Likewise, If the truth value of an existentially quantified
formula $(\exists x.\ F)$ is \enquote{true}, only one instance $F[x\mapsto a]$
is displayed, where $a$ is the first value encountered in the evaluation that
makes $F$ \enquote{true}. By inspecting such an evaluation tree, we may quickly
get the essential information to understand how the (possibly unexpected) truth
value of a formula (e.g., of an invalid verification condition) was derived.

The scalability of the RISCAL visualization tools is currently limited: if
the visualized graph (execution trace or evaluation tree) exceeds the actual
area of the window, only a part of the graph is displayed and scrollbars are
shown that allow the user to navigate the displayed part within the graph;
this is because the Eclipse GEF4 Zest software underlying the RISCAL
visualization does not support zooming. There are also limits on the size of
the evaluation tree for which the algorithms used by this software can
compute a layout in a reasonable amount of time; thus RISCAL currently only
visualizes trees with a configurable maximum (currently: 500) of nodes on
each layer of abstraction.

\subsection{Web-Based Exercises}
\label{sect:web}

\begin{figure}
\begin{center}
\includegraphics[width=0.7\textwidth]{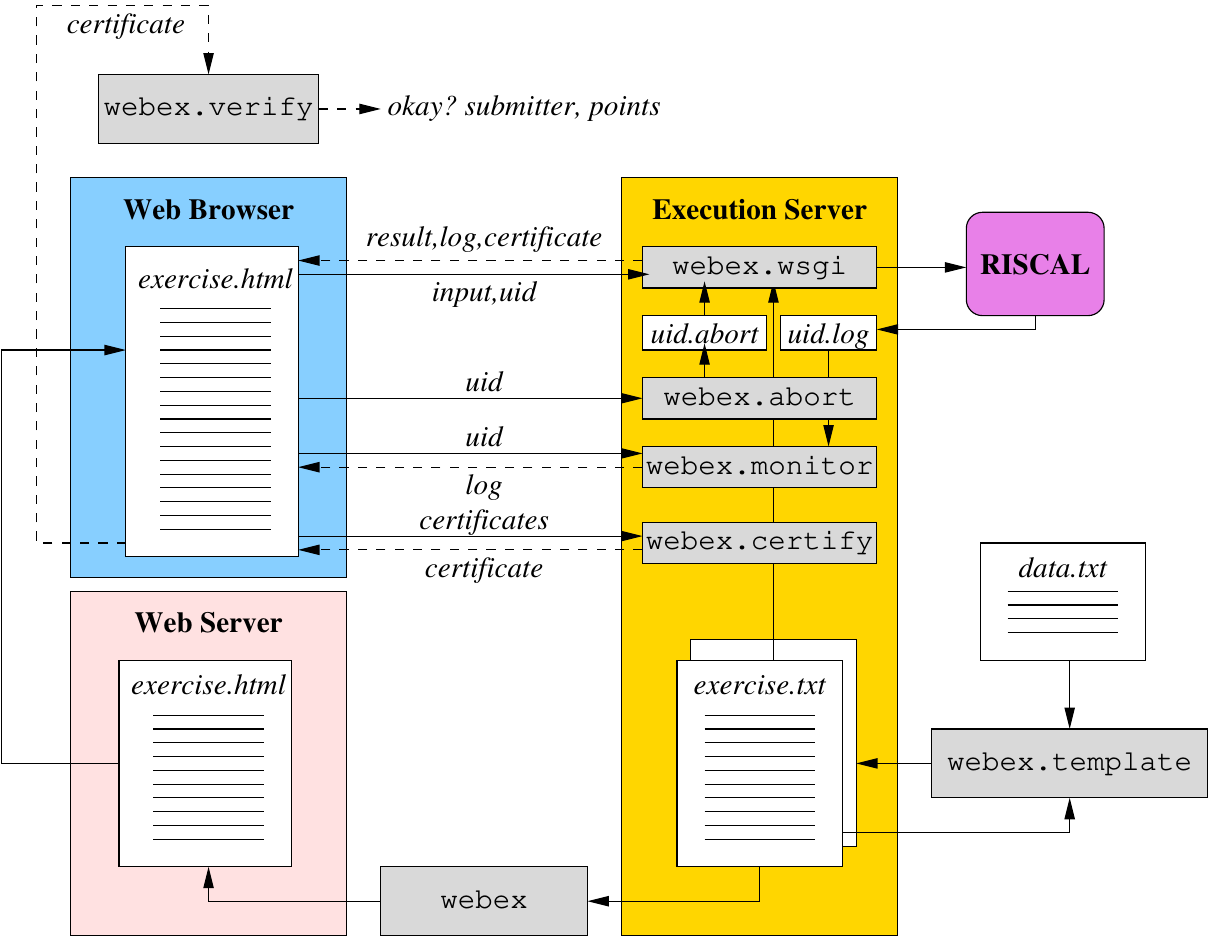}
\end{center}
\caption{The RISCAL WebEx Software Architecture}
\label{fig-software}
\end{figure}

As will be discussed in the following sections, the use of RISCAL in
educational scenarios has so far required the installation of software on
every student's own computer, either a virtualization software (for the
execution of a downloadable virtual machine that has RISCAL preinstalled) or
a remote desktop solution (for the execution of RISCAL on a remote server).
However, especially in undergraduate education it would be advisable to
provide access to RISCAL also via the web without requiring any local
software installation. Furthermore, the elaboration of an exercise should
not necessarily expose a complete RISCAL specification but only those parts
that are relevant for the student (hiding in particular those parts that
describe how the correctness of the exercise is checked). For this purpose,
we have recently developed the \enquote{RISCAL WebEx}
software~\cite{Schreiner2018e} (not to be confused with the CISCO WebEx
video conferencing software), an experimental Python-based software
framework that allows to turn with moderate effort annotated RISCAL
specifications into web-based exercises.

Figure~\ref{fig-software} displays the architecture of this software and the
workflow of its use. In a nutshell, the command \texttt{webex} generates
from an annotated RISCAL specification file \emph{exercise.txt} (that is
deployed on the execution server) an HTML file \emph{exercise.html} that
represents an exercise form; this file is deployed on the web server. The
client downloads \emph{exercise.html} from the web server to her web
browser. She enters her name into the exercise form and performs some task
described in the form by providing some input and then triggering some
RISCAL action to check that input. The execution service \texttt{webex.wsgi}
runs on the execution server, a network server that directly communicates
with the user's web browser. This execution service produces from
\emph{exercise.txt} and the user input a plain RISCAL file and starts a
RISCAL process to execute the action denoted by the user on that file. When
the RISCAL process terminates (normally, by user abortion, or by timeout),
\texttt{webex.wsgi} returns the result status of the RISCAL process (success
or failure) together with the produced output and a digital certificate of
the execution (signed with a private key of the execution server) to the web
browser; the certificate includes the user name, the action performed, and
the grade points earned. Ultimately, from the certificates of all tasks that
the user has performed, a summary certificate is produced that the user may
store in a file and submit as the result of the exercise. The lecturer may
use a script \texttt{webex.verify} (which needs access to the private server
key) to check the authenticity of the certificate and to retrieve the number
of grade points that she has earned. Furthermore, an existing exercise file
can be instantiated by a script \texttt{webex.template} with data from
another file, thus producing a collection of exercises.

\begin{figure}[t]
\begin{center}
\includegraphics[width=0.93\textwidth]{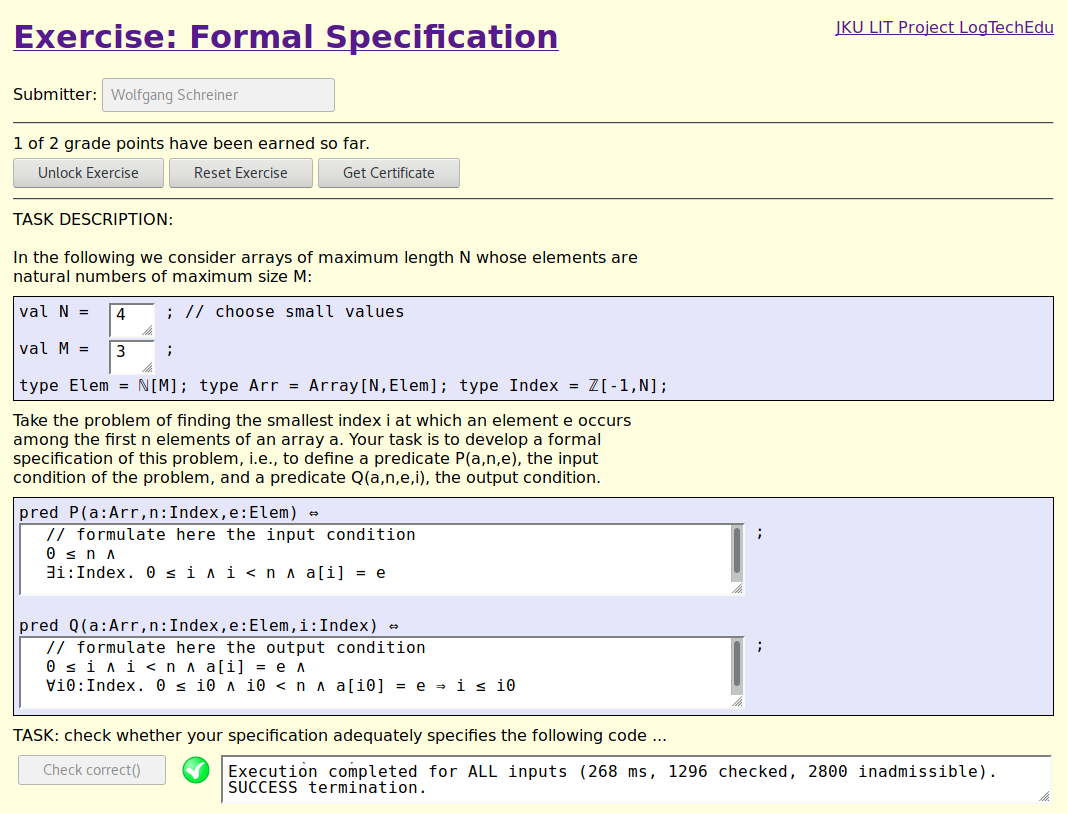}
\end{center}
\caption{A RISCAL Web Exercise}
\label{fig-webex}
\end{figure}

Figure~\ref{fig-webex} displays an example of such a web exercise
generated from an annotated RISCAL file whose content is sketched below:

\begin{xverbatim}
//@ <webex id="firstOrderPragmatics" type="riscal" header="Exercise: Formal Specification">

/*@ <display>
TASK DESCRIPTION: 

In the following we consider arrays of maximum length N whose elements are
natural numbers of maximum size M: 
</display> @*/

//@ <public>
val N = /*@<input id="N" cols="2">@*/4/*@</input>@*/; // choose small values
val M = /*@<input id="M" cols="2">@*/3/*@</input>@*/;
type Elem = ℕ[M]; type Arr = Array[N,Elem]; type Index = ℤ[-1,N];

//@ <private>
...
/*@ <button id="correct" label="Check correct()" action="correct">
      <argument value="-opt-si"> <argument value="0">
</button>@*/
\end{xverbatim}

This exercise combines displayed text with formal content (some of which is
hidden from the student) and input from the student. After having provided the
requested input, the student presses the button upon which the RISCAL software
on the remote server is invoked to check its correctness; the success status and
the output of the check (for further investigation in case of errors) are
displayed by a success icon respectively in a (resizable) output field.

The RISCAL WebEx software is still in an experimental stage; more details on its
use (including the exact grammar of the annotation language) are given
in its manual~\cite{Schreiner2018e}; the RISCAL web page~\cite{RISCAL}
lists more examples of RISCAL-based web exercises.

\section{The Application of RISCAL in Education}
\label{sect:education}

In this section we are going to report on applications of RISCAL in educational
scenarios, past, current, and future (respectively envisioned) ones.

\subsection{A Course on Formal Methods}
\label{sect:formal}

The main use of RISCAL so far has been in an annual course on \enquote{Formal
Methods in Software Development} at the Johannes Kepler University (JKU) Linz
for master students in computer science and computer mathematics. This course
deals with the formal specification and verification (respectively
falsification) of computer programs and is organized in three parts: the first
part deals with fundamental concepts and calculi (the Hoare calculus, Dijkstra's
predicate transformer calculus, a relational calculus), the second part with the
application of these techniques to actual Java programs, the third one with
concurrency and nondeterminism. All parts are accompanied by the demonstration
of software tools and their use by students to elaborate ten home assignments.
The student groups are of moderate size (about 25~participants) and have
adequate technical background to install and operate on their own computers a
virtual machine with RISCAL preinstalled.

Before the academic year 2017/2018, we used in the first part of this course
(with four home assignments) the RISC ProofNavigator~\cite{Schreiner2008} as an
interactive proving assistant on top of which the RISC
ProgramExplorer~\cite{Schreiner2012} provided an integrated specification and
verification environment for a subset of Java called \enquote{MiniJava}. With
these tools, the adequacy of specifications and the correctness of programs with
respect to these specifications could in essence only be judged by proving the
validity of (manually derived respectively automatically generated) verification
conditions; if an attempt to such a proof failed, it remained often unclear to
students whether this was due to a poor proof strategy or inadequate formal
specifications respectively annotations (to at least rule out actual program
errors, correct programs were handed out). Furthermore, sometimes a verification
succeeded only because the student's specification was inadequate (e.g., a
postcondition was trivially satisfied due to an error in the logical formulation).

In the winter semester 2017/2018, we complemented the first four assignments by
the use of RISCAL version 1.x. The topic of Assignment~1
was the development of formal specifications and the analysis of their adequacy
in various ways~\cite{Schreiner2018a}: 
\begin{itemize}
\item by executing functions that were implicitly defined from the
pre- and postconditions (illustrating the input/output relationship
	expressed by the specification); 
\item by checking
whether preconditions and postconditions are satisfiable but non-trivial (ruling
	out logical errors that make the specification trivial); 
\item by checking
whether postconditions define the results uniquely (if intended);
\item by checking whether every output of a given procedure that implements
the specification indeed satisfies the output condition.
\end{itemize}
For these checks students had to formulate the respective functions
and theorems manually. Assignment~2 dealt with formal verification and the
corresponding annotation with loops by invariants; students first had to
validate specifications and invariants (by checking their validity in all
possible executions of RISCAL procedures) and then to manually derive the
verification conditions and check their validity with RISCAL (which implies that
the invariants are sufficiently strong). In Assignments~3 and~4, specifications
and invariants were first validated as in Assignment~2 (without generating
actual verification conditions) and then transferred to a corresponding MiniJava
program; the verifications conditions were generated by the RISC
ProgramExplorer and interactively verified with the RISC ProofNavigator. The use
of RISCAL in the later  assignments ensured that proofs were only attempted
after validation had ensured that specifications and annotations were
not too strong (but invariants might be still too weak).

In the winter semester 2018/2019, RISCAL version 2.x enabled us to augment
the assignments: First, in all assignments the visualization of formula
evaluation trees could be used to better understand how the truth values of
formulas were derived; furthermore, in Assignments~2--4 the conditions for
specification validation conditions and for program verification were
automatically generated from the annotated programs, such that in
Assignments~3 and~4, the specifications and annotations could be analyzed
more thoroughly than in the previous year.

A final anonymous evaluation of the course software by the participants (13
reports were returned from a total of 21 students graded) indeed indicated a
very high satisfaction with the use of RISCAL both from the point of ease of
use (average rating: 4.2 of 5) and learning success (average rating: 4.3 of
5); the ratings were significantly better than for our previously developed
RISC ProgramExplorer/ProofNavigator (ease of use: 3.5, learning success:
3.5) and summarily higher than for any other of the six toolsets used in the
course (the second most popular one was ESC/Java2 with ratings of 4.2 and
3.8, respectively). As for the actual effect on exercise grading, however,
we have to report a null result: as in the previous two incarnations of the
course the average grades were in the mostly range of 80--95 points out of
100 with no statistically significant effect visible by the use of RISCAL.

Furthermore, a recently completed bachelor thesis~\cite{Payr2018} has
elaborated the formal specification and verification of fundamental searching
and sorting algorithms (including the major asymptotically fast ones) on
sequences in various representations (arrays, recursive lists, pointer-linked
lists); here the student was able to elaborate on his own annotations (loop
invariants) that are sufficiently strong to generate verification conditions
that pass all RISCAL checks in finite models. This is to us a strong indication
that the use of RISCAL may be indeed able to quickly enhance a student's
understanding of the formalization of theories and algorithms.

\subsection{A Course on Logic}
\label{sect:logic}

Since 2013 we have collaborated with three colleagues in teaching JKU's
introductory \enquote{Logic} course for (about 200) first semester
undergraduate students in computer science. This course consists of three
modules on propositional logic/satisfiability solving (SAT), first order
predicate logic (FO), and satisfiability module theories (SMT). Having
started with a traditional \enquote{paper and pencil} course, we have over
the years incorporated the use of various logic-based software tools: a SAT
solver (Limboole), an interactive proving assistant for predicate logic (the
RISC ProofNavigator), an automated reasoning system (Theorema), and SMT
solvers (Boolector, Z3). However, we have refrained from making the use of
these tools mandatory, because they must be installed on the students' own
computers and we could not provide sufficient technical support for a large
number of course participants (all beginners with varying technical
background), if they were obliged to use the software. Until the academic
year 2017/2018, we therefore confined the use of these tools to four
voluntary \enquote{laboratory assignments} in which students had the
possibility to earn additional grade points by solving specific problems
with these tools. Since these assignments were outside the main stream of
the course, only a minor fraction of the students (about 5--10\%) took this
opportunity, either because of special interest in the topic or just because
of an urgent need for extra grade points to pass the course. While we thus
limited our support efforts, we also artificially constrained the potential
educational effects of the use of logic software.

However, in the current winter semester 2018/2019, we have experimentally
integrated into the course an additional element to foster the more wide-spread
use of these tools (still without making it compulsory): every week we give
voluntary \enquote{bonus assignments} (twelve in total) by which a student may
earn up to 20\% of the grade points that are awarded by the weekly written tests
on which the overall grading of the course is mainly based. These assignments
are of a moderate complexity (much smaller than that of the laboratory
assignments), are to be performed by the students at home, and are to be
submitted before the corresponding tests; furthermore the assignments are
designed such that the correctness of results can be essentially self-checked by
the students via the utilized software. Since thus mostly correct solutions are
submitted and since also only a minor part of the grading is based on these home
assignments (typically sufficient to improve the grade by one level),
essentially only the completeness and plausibility of solutions is judged by
teaching assistants. To mitigate the effect of cheating, we include some element
of \enquote{personalization} to the assignment which ranges from the automatized
generation of individualized problem inputs for every student to simply
requiring from every student to submit a screenshot that demonstrates her
personal use of the software.

In this ongoing effort, RISCAL is being used for one of the laboratory
assignments (replacing the previous use of the RISC ProofNavigator) and for
three of the bonus assignments. All these assignments are associated to the
first half of the module on first order predicate logic dealing with its syntax,
semantics, and pragmatics (the other half of this module mainly deals with
proving for which the Theorema system is used). In the current semester, the use
of RISCAL is still based on a virtual machine installation and on a remote server
installation (the first prototype of the web-based framework for RISCAL
exercises described in Section~\ref{sect:web} has been just recently completed
and will be used from the next academic year on). Thus each RISCAL assignment
currently consists of one or more RISCAL skeleton specifications where the
students have to complete some definitions and then execute some operations to
check the correctness of their solutions:

\begin{itemize} 

\item In the \enquote{syntax} assignment, students have to parse given predicate
logic formulas and include parentheses to make their structure unique; it is the
logic equivalence to the original formula that is checked. Furthermore, students
have to translate informal statements to logic formulas; here the logic
equivalence to another formalization is checked.

\item In the \enquote{semantics} assignment, students have to determine the
satisfying assignments of given formulas (respectively determine the truth
values of closed formulas); operations are given that not only check the
correctness of the solutions but also give hints (such as the number of expected
solutions or the truth values arising from all possible assignments).
Furthermore, students have to decide whether one formula is a logical
consequence of (respectively logically equivalent to) another one and to
transform a formula into a logical equivalent form with a certain syntactic
property; again the results can be checked (at least validated) by executing
some given operations.

\item In the \enquote{pragmatics} assignment, students have to translate several
given informal problem specifications into logic formulas that capture the input
respectively output conditions of the problems; this partially also involves the
definition of auxiliary functions and predicates. The results can be validated
by executing some operations as described in Section~\ref{sect:logic}.

\end{itemize} 
As the results of the RISCAL assignments, students submit the completed
specification file (together with a screenshot demonstrating the personal use of
the software).

Concerning the actual effects on the use of RISCAL and other software used in
this course we can report both on an anonymous student evaluation (we
received evaluations from 133 students) and the actual outcomes of the tests
(197 students were graded, 24 students dropped out after the first or second
lecture without receiving a grade):

\begin{itemize}

\item The question \enquote{Which software was helpful?} was answered by 82
students with \enquote{RISCAL}, clearly trailing \enquote{Limboole} (named
121 times) but significantly ahead of the other two tools (which were named
51 and 54 times); so among those tools applicable to richer logics than pure
propositional logic, RISCAL was the most popular one. Also there was only a
very small number of problems reported with the virtual machine (5),
documentation (3), and ease of use (6).

\item As for the question \enquote{Why did you attempt the bonus
assignments?} (no classification with respect to the specific tool), only a
disappointing number of 37 answered by \enquote{interest} or the potential
of better \enquote{understanding}, while 77 reported as their core reason
\enquote{easiness} or the potential to earn additional \enquote{points}.
Furthermore, the question \enquote{Would you like the introduced software to
be part of your other courses in the curriculum?} was answered positively
only by 41 students with \enquote{Limboole} mentioned 25 times and RISCAL
mentioned 12 times (the other two tools were mentioned 10 and 11 times).
Interestingly, however, the most positive impact on interest in the course
was attributed to \enquote{Software} (named 18 times), while the most
positive impact on actually understanding was attributed to
\enquote{Exercises} (named 36 times) and \enquote{Moodle Quizzes} (named 26
times) much more than \enquote{Software} (named 13 times).

\item As for the actual impact on grading, we report subsequently in detail
only on the use of RISCAL. 85 out of 191 participating students attempted at
least one of the three RISCAL-related bonus exercises (each bonus exercise
was performed by 76, 74, and 54 students, respectively). Since these are
many more students than those who did a voluntary RISCAL-related laboratory
exercise~(9), our goal to spread the use of logic software tools among
students was thus apparently achieved. The majority of the submissions
indeed earned the students the complete bonus of one point per bonus
assignment; since, however, the maximum of classroom exercise and bonus
exercise could not exceed 5 points, in average only 1.48 extra points per
student actually contributed to the grading.

\item There was a strong correlation between work on bonus assignments and
performance in classroom assignments: out of 85 students with all three
classroom assignments positive also 56 students did some bonus assignment;
however, out of 54 students with only zero or one positive classroom
assignments only 7 did some bonus assignment (which actually enabled 3 of
them to pass the threshold of two positive assignments). Interestingly, also
of the 9 students who did the lab exercise, 5 students had already 3
positive classroom assignments (but apparently wanted to improve their
grade) while 1 student had 1 such assignment (and passed by the laboratory
the threshold of 2 positive assignments). This may indicate that the
performance of students improves by the use of the tool but also that only
students with a higher performance use the tools (probably both are true).
Thus, while a strong correlation is visible, the direction of cause and
effect is not clear.

\item As for the overall effect on grading, a comparison between the current
invocation of the module and the corresponding instances in the previous
year is difficult, since some of the assignments have considerably changed.
All in all a significant difference in the overall outcome of the
RISC-related part of the first-order logic module is not discernible (the
overall number of about 3.4 points earned in the classroom exercises, not
considering the bonus exercises, remained almost the same).

\end{itemize}

In a nutshell, our overall experience is that indeed a large number of
students attempted the RISCAL bonus exercises (and performed well there) and
that those who did so achieved significantly better results in the classroom
exercises than those that did not. However it also seems that really weak
students (subsequently failing the course) mostly ignored the software and
thus also did not make use of extra chances to pass the course. Furthermore,
the final evaluation indicates that the main motivation of many students to
use the software was to earn additional points rather than an intrinsic
interest or the desire to improve their understanding; nevertheless
\enquote{Software} was listed as the highest impact factor for raising their
interest (while it was not named as a significant factor for improving their
understanding).

%Nevertheless, in the whole course, mainly by the extra bonus points, also
%the average grade has improved from 3.6 in 2017 to 3.2 in 2018. While the rate
%of negative grades was reduced from 36\% to 28\%, the number of early
%dropouts increased from 6\% to 11\%. All of this, however, is in the
%realm of statistical fluctuations.

\subsection{Courses on Modeling and Programming}
\label{sect:future}

In the forth-coming summer semester 2019, we will start at JKU (as a
collaborative effort of several lecturers) a new course on \enquote{Formal
Modeling} for undergraduate students in mathematics. Approximately one third of
this course will deal with the modeling of computational problems respectively
computational systems in first order logic; here we will apply RISCAL to check
the validity of the models respectively to illustrate their executable aspects.
This exposition will complement other already existing courses in the degree
program that mainly focus on modeling continuous domains in science and
engineering with techniques from analysis and numerical mathematics.

The audience/level of this course will differ from those of the previously described
courses on \enquote{Logic} (for first semester students in computer science) and
\enquote{Formal Methods in Software Development} (for master students in
computer science and computer mathematics) by addressing students of mathematics
that are in the later phase of their bachelor program. Likewise the content
of this course differs from that on \enquote{Logic} (focusing on the basics of
first order predicate logic) and \enquote{Formal Methods} (focusing on program
specification and verification) by emphasizing the applicability of logic
modeling to problems in discrete mathematics, computer algebra, computational
logic, etc. The goal is to demonstrate also to mathematicians the applicability
and practical importance of logic-based modeling techniques which allows one
also to draw definite (precise and formally justifiable) conclusions about
certain problem domains (rather than just computing numerically approximated
solutions).

As a longer-term vision we hope to achieve some impact also on the basic
education of computer science students in the actual core of their field, the
design and implementation of computer programs. Corresponding introductory
courses on \enquote{Software Development} or \enquote{Algorithms and Data
Structures} typically present the field on the basis of examples (which students
mimic/adapt/generalize to develop their own solutions to given problems)
respectively of general recipes whose propriety (domain of applicability and
correctness) however is again explained by examples; consequently students are
trained to judge the adequacy of their own developments by example-based
\enquote{trial and error} (also known as \enquote{testing}). Thus in our own
experience even master students of computer science are hard-pressed when given
the task to develop from a simple informal problem specification a code snippet
of twenty lines that solves the problem in a completely correct way
(considering also all special, boundary, respectively corner cases). It is hard
to imagine how on this basis complex software systems can be developed that are
expected to be reliable, safe, and secure.

A possible approach to improve this situation is to incorporate into the
education on algorithm respectively software development also assignments where
the correctness of the solution is automatically checked on the basis of
logic-based software. For example, as sketched in Section~\ref{sect:web}, a
RISCAL-based web assignment may
\begin{itemize}
\item display an informal description of a computational problem to be
solved,
\item provide the public skeleton of a procedure with an embedded area where 
the student may enter the code of their implementation to solve this problem,
\item contain a hidden annotation of the code by a formal specification
of its pre- and post-condition,
\item expose a button to execute the procedure for all possible inputs.  
\end{itemize}
The student may then self-check the correctness of her solution: if the
solution fails for some input argument (i.e., the code aborts or produces a wrong
result), the student is correspondingly informed from the RISCAL execution
output and may correct her solution. Since solutions have been self-checked, it
can be expected that mostly correct solutions will be submitted; furthermore,
the results can be by this logic-based execution framework automatically graded
with respect to its formal contract (extending the functionality of conventional
autograder software~\cite{VPL2015} which automatically grades programs
by testing~\cite{Edwards2008} or static code analysis~\cite{Yulianto2014}).

While we are not responsible for any course that is suitable to embody the vision
sketched above, we will approach some lecturers responsible for the beginners'
education in algorithm and software development at JKU; we hope to convince them
by a demonstration of the capabilities of RISCAL (with appropriately prepared
examples and lecturing materials) to experimentally include some of these
envisioned elements into their own courses.

\section{Conclusions}
\label{sect:conclusions}

Compared with the status reported in~\cite{Schreiner2018a}, RISCAL has
significantly matured by incorporating additional features that are relevant
its application in education. Most important, however, we have started to
gain actual experience with the use of the software for educational
scenarios, in particular in graduate classes on formal methods and logic and
in the elaboration of two bachelor theses on formal
specification~\cite{Brunhuemer2017} and verification~\cite{Payr2018}. This
first experience not only shows that the software is stable and (compared to
other software we have used before) easy to use; it also shows that students
are really able to develop formal material in a faster way and of
substantially higher quality than with the proof-based tools we have used
before. Actually, it is the first time we had the feeling that (some)
students \emph{enjoyed} using the software due to its immediate feedback on
the meaning and adequacy of the formalisms they develop.

As for our initial experience with the (optional) use of RISCAL in an
undergraduate introductory course on logic, we also experienced that
students had little technical problems with RISCAL. Furthermore, there was a
strong correlation visible between the use of the software by students and
their subsequent performance in traditional classroom exercises. However,
whether the software indeed improved their understanding or whether just the
stronger students used the software remains unclear; students that performed
poorly in the classroom assignments generally ignored the software. A
subsequent evaluation revealed that students were mainly motivated to use the
software for improving their grades; nevertheless the factor
\enquote{Software} was most often named as a factor to raise interest in the
presented topics. This experience may indicate that the use of RISCAL-like
tools in education may achieve true profit once the students already know
enough to appreciate the potential of such tools; in the initial stages of
their education, they may serve more as additional factors to raise
subsequent interest in the presented topics.

However, all this is still mainly hypothesis based on anecdotal evidence. In
addition to enhancing RISCAL further (e.g., by applying also SMT based
checking techniques) and to applying RISCAL in more courses (e.g., on formal
modeling mathematical domains), our main next goal is to evaluate the
educational effect of the use of this tool in a systematic way and provide
statistically sound scientific evidence of its effects. The core
difficulty will be to design a suitable experimental scenario and integrate
it into courses such that on the one side actual scientific evidence is
gathered (we have to avoid, e.g., self selection bias) and on the other hand
the actual education is not compromised (we cannot force students into an
experiment with potential detrimental effects on their educational outcome),
also considering ethical problems (how can we e.g. justify to place students
into a control group that does not use the software even if we have already
strong evidence that it improves their study results). Here more and deeper
investigations are required.

\bibliographystyle{eptcs}
\bibliography{main}

\end{document}

%% file: unicodechars.tex
\newunicodechar{→}{\ensuremath{\rightarrow}}
\newunicodechar{≔}{\ensuremath{:=}}
\newunicodechar{×}{\ensuremath{\times}}
\newunicodechar{⊆}{\ensuremath{\subseteq}}
\newunicodechar{⇔}{\ensuremath{\Leftrightarrow}}
\newunicodechar{¬}{\ensuremath{\neg}}
\newunicodechar{∧}{\ensuremath{\wedge}}
\newunicodechar{∨}{\ensuremath{\vee}}
\newunicodechar{⇒}{\ensuremath{\Rightarrow}}
\newunicodechar{≤}{\ensuremath{\leq}}
\newunicodechar{≥}{\ensuremath{\geq}}
\newunicodechar{∞}{\ensuremath{\infty}}
\newunicodechar{∈}{\ensuremath{\in}}
\newunicodechar{∪}{\ensuremath{\cup}}
\newunicodechar{∩}{\ensuremath{\cap}}
\newunicodechar{⋃}{\ensuremath{\bigcup}}
\newunicodechar{⋂}{\ensuremath{\bigcap}}
\newunicodechar{∅}{\ensuremath{\emptyset}}
\newunicodechar{≠}{\ensuremath{\neq}}
\newunicodechar{∉}{\ensuremath{\not\in}}
\newunicodechar{∊}{\ensuremath{\in}}
\newunicodechar{⊢}{\ensuremath{\vdash}}
\newunicodechar{ε}{\ensuremath{\epsilon}}
\newunicodechar{⌚}{\ensuremath{\clock}}
\newunicodechar{⦇}{\ensuremath{\llparenthesis}}
\newunicodechar{⦈}{\ensuremath{\rrparenthesis}}
\newunicodechar{∼}{\~{}}
\newunicodechar{ℕ}{\ensuremath{\mathbb{N}}}
\newunicodechar{ℤ}{\ensuremath{\mathbb{Z}}}
\newunicodechar{ℙ}{\ensuremath{\mathbb{P}}}
\newunicodechar{⊤}{\ensuremath{\top}}
\newunicodechar{⊥}{\ensuremath{\bot}}
\newunicodechar{⋅}{\ensuremath{\cdot}}
\newunicodechar{〈}{\ensuremath{\langle}}
\newunicodechar{⟨}{\ensuremath{\langle}}
\newunicodechar{〈}{\ensuremath{\langle}}
\newunicodechar{〉}{\ensuremath{\rangle}}
\newunicodechar{⟩}{\ensuremath{\rangle}}
\newunicodechar{〉}{\ensuremath{\rangle}}
\newunicodechar{∑}{\ensuremath{\sum}}
\newunicodechar{Σ}{\ensuremath{\Sigma}}
\newunicodechar{∏}{\ensuremath{\prod}}
\newunicodechar{Π}{\ensuremath{\Pi}}
\newunicodechar{∀}{\ensuremath{\forall}}
\newunicodechar{∃}{\ensuremath{\exists}}

%% file: commands.tex
\newcommand{\button}[1]{\raisebox{-0.1\baselineskip}
{\includegraphics[height=0.9\baselineskip]{#1.png}}}

\newenvironment{xverbatim}{\quote\verbatim}{\endverbatim\endquote}